# Electrical and optical control of single spins integrated in scalable semiconductor devices

Christopher P. Anderson[1,2,*], Alexandre Bourassa[1,*], Kevin C. Miao[1], Gary Wolfowicz[1], Peter J. Mintun[1], Alexander L. Crook[1,2], Hiroshi Abe[3], Jawad Ul Hassan[4], Nguyen T. Son[4], Takeshi Ohshima[3], David D. Awschalom[1,2,5,†]

[1]*Pritzker School of Molecular Engineering, University of Chicago, Chicago, Illinois 60637, USA*

[2]*Department of Physics, University of Chicago, Chicago, Illinois 60637, USA*

[3]*National Institutes for Quantum and Radiological Science and Technology, 1233 Watanuki, Takasaki, Gunma 370-1292, Japan*

[4]*Department of Physics, Chemistry and Biology, Linköping University, SE-581 83 Linköping, Sweden*

[5]*Center for Molecular Engineering and Materials Science Division, Argonne National Laboratory, Lemont, IL 60439, USA*

*\* These authors contributed equally to this work.*
*† email:awsch@uchicago.edu*

**Spin defects in silicon carbide have exceptional electron spin coherence with a near-infrared spin-photon interface in a material amenable to modern semiconductor fabrication. Leveraging these advantages, we successfully integrate highly coherent single neutral divacancy spins in commercially available p-i-n structures and fabricate diodes to modulate the local electrical environment of the defects. These devices enable deterministic charge state control and broad Stark shift tuning exceeding 850 GHz. Surprisingly, we show that charge depletion results in a narrowing of the optical linewidths by over 50 fold, approaching the lifetime limit. These results demonstrate a method for mitigating the ubiquitous problem of spectral diffusion in solid-state emitters by engineering the electrical environment while utilizing classical semiconductor devices to control scalable spin-based quantum systems.**

**Introduction**

Solid-state defects have enabled many proof-of-principle quantum technologies in quantum sensing(*1*), computation(*2*) and communications(*3*). These defects exhibit atom-like transitions that have been used to generate spin-photon entanglement and high-fidelity single-shot readout(*4*) enabling demonstrations of long-distance quantum teleportation, entanglement distillation and loophole-free tests of Bell's inequalities(*3*).

However, fluctuating electric fields and uncontrolled charge dynamics have limited many of these technologies(*1, 4–7*). For example, charge stability and photon indistinguishability are major problems that reduce entanglement rates and fidelities in quantum communication experiments(*4–6*). In particular, indistinguishable and spectrally narrow photon emission is required in order to achieve high-contrast Hong-Ou-Mandel interference(*8*). This indistinguishability has been achieved with some quantum emitters through dc Stark tuning the optical lines into mutual resonance(*9, 10*). Meanwhile, a variety of strategies(*1, 6, 11–13*) have been proposed to reduce spectral diffusion(*14*) and blinking(*15*), but consistently achieving narrow and photostable spectral lines remains an outstanding challenge(*16*). In addition, studies of charge dynamics(*17, 18*) have enabled quantum sensing improvements(*1, 7*) and spin-to-charge conversion(*19*) allowing electrical readout of single spin defects(*20*). However, these experiments have largely been realized in materials such as diamond where scalable nanofabrication and doping techniques are difficult to achieve.

In contrast, the neutral divacancy ($VV^0$) defect in silicon carbide (SiC) presents itself as a candidate spin qubit in a technologically mature host, allowing for flexible fabrication, doping control, and availability on the wafer scale. These defects display many attractive properties including all-optical spin initialization and readout(*21*), long coherence times(*22*), nuclear spin control(*23*), as well as a near-infrared high-fidelity spin-photon interface(*24*). However, $VV^0$ have suffered from relatively broad optical lines(*24*), charge instability(*18*) and relatively small Stark shifts(*10*). Furthermore, the promise of integration into classical semiconducting devices remains largely unexplored.

Here, we utilize the mature semiconductor technology that SiC provides to create a p-i-n structure that allows tuning of the electric field and charge environment of the defect. First, we isolate and perform high fidelity control on highly coherent single spins in the device. We then show that these devices enable wide dc Stark tuning while maintaining defect symmetry. Interestingly, we also demonstrate that charge depletion in the device mitigates spectral diffusion thus greatly narrowing the linewidths in the optical fine structure. Finally, we use this device as a testbed to study the photoionization dynamics of single $VV^0$, resulting in a method for deterministic optical control of the defect charge state.

The effects presented here suggest that doped SiC structures are flexible and scalable quantum platforms hosting long-lived single spin qubits with an electrically-tunable high-quality optical interface. The demonstrated reduction in electric field noise may lead to increased spin coherence(*25*), electrical tuning of 'dark' spins in quantum sensing(*26*), while charge control may potentially extend the memory time of nuclear spins(*27*). Additionally, this platform opens unique avenues for spin-to-charge conversion, electrically-driven single photon emission(*28*), electrical control(*29*) and readout(*20*, *30*, *31*) of single spins in SiC semiconductor devices.

**Isolated single defects in a semiconductor device**

We first isolate and control single $VV^0$ in a SiC p-i-n diode created through commercial growth of doped SiC epilayers. After growth, electron irradiation and subsequent annealing creates single, isolated $VV^0$ defects. We fabricate microwave striplines and Ohmic contact pads allowing for spin manipulation and electrical gating (Fig. 1A) (*32*). In contrast to other defects in SiC such as the isolated silicon vacancy(*33*), the divacancy is stable above $1600\,^\circ$C (*34*) making it compatible with device processing and high temperature annealing to form Ohmic contacts.

Spatial photoluminescence (PL) scans of the device show isolated emitters corresponding to single $VV^0$ (Fig. 1B), as confirmed by second-order correlation ($g^{(2)}$) measurements (Fig. 1B, inset)(*32*). The location in depth of the observed defects is consistent with isolation to the i-type layer, as expected from formation energy calculations(*35*) and the local Fermi level(*36*). This depth localization provides an alternative to delta-doping(*37*), which is not possible with intrinsic defects, facilitating positioning and control in fabricated devices (Fig. S1). Additionally, due to the diode's highly rectifying behavior at low temperature, large reverse biases are possible with low current (Fig. 1c)(*32*).

Sweeping the frequency of a narrow-line laser, we obtain photoluminescence excitation (PLE) spectra of the optical fine structure of these single defects (Fig. 1D). Using the observed transitions for resonant readout and preparation, we perform high-contrast Rabi oscillations of isolated VV$^0$ in the p-i-n structure (Fig. 1E)(*32*). The contrast exceeds 98%, improving on previous demonstrations through the use of resonant spin polarization(*24*). Additionally, a single spin Hahn-echo decay time of 1.0±0.1 ms is measured for spins in the device (Fig. 1F), consistent with previous ensemble measurements(*22*). The long Hahn-echo times and high-fidelity control demonstrate that integration into semiconductor structures does not degrade the spin properties of VV$^0$. This isolation and control of highly coherent spin qubits achieved in these functioning semiconductor devices unlocks the potential for integration with a wide range of classical electronic technologies.

**Large Stark shifts in a p-i-n diode**

Since the (*hh*) and (*kk*) divacancies(*32*) in SiC are nominally symmetric along the c-axis (growth axis), the geometry of the diode allows for large electric fields which mostly conserve the symmetry of the defect. Therefore, wide tuning of the VV$^0$ optical structure is possible, while reducing unwanted mixing from transverse or symmetry-breaking components of the excited state Hamiltonian(*9*, *24*, *38*). Because the i-type region can be relatively thin (10 µm here), the applied voltage is dropped over a much smaller region than if a bulk sample were used(*10*), leading to significantly larger Stark shifts for a given applied voltage. In principle, this region can be reduced to a thickness that exceeds limitations from optical access with metal planar gates (limited by the optical spot size of ~1 µm). Furthermore, it is possible to use doped layers as *in-situ* transparent native contacts to Stark tune and control localized defects in suspended photonic or phononic structures(*39*) enabling complex hybrid electrical, photonic and phononic devices.

In our p-i-n junction device, we apply up to -420 V in reverse bias. Our results show Stark tuning of several hundreds of GHz on different defects of the same type and on inequivalent lattice sites where the Stark shift is between 0.4-3.5 GHz/V after a threshold is passed (Fig. 2A). For example, we observe a (*hh*) divacancy shifted by more than 850 GHz (2.5 meV) at -420 V and a (*kh*) divacancy shifted by more than 760 GHz at -210 V (Fig. 2B). These shifts are among the largest reported for any single spin defect to date and were only limited by the voltage output of our source. We expect that due to the high dielectric breakdown field of SiC, even higher shifts of a few THz are possible. The high-field limit of these shifts correspond to estimated dipole moments (d$_\parallel$) of 11 GHz m/MV and 4.5 GHz m/MV for (*hh*) and (*kk*) divacancies respectively, consistent with previous reports(*10*, *40*). For the (*kh*) basal divacancy observed, the estimated transverse dipole moment is around d⊥ = 35 GHz m/MV. Furthermore, since the Stark shift represents a measure of the local electric field, we conclude that negligible field is applied to the $VV^0$ before a certain threshold voltage where the depletion region reaches the defect(*41*). This results from non-uniform electric fields in the diode caused by residual n-type dopants in the intrinsic region (Fig. 2C, see supplemental text).

Overall, our system could be used as a widely frequency-tunable, spectrally narrow source of single photons. To our knowledge, this represents one of the highest Stark shift to linewidth ratios (>40,000) obtained in any single photon source (Table S1). These characteristics make our system ideally suited for tuning remote defects into mutual resonance and for frequency multiplexing of entanglement channels(*42*). Interestingly, the tunability range is so wide that it could even enable the tuning of a (*hh*) divacancy into resonance with a (*kk*) divacancy, allowing for interference and entanglement between different species of defects. This wide tunability stems from the rectification behavior of the diode which allows large electric fields without driving appreciable currents, which can degrade spin and optical properties. Furthermore, the observed sensitivity of the optical structure of single $VV^0$ defects could serve as a nanoscale electric field sensor enabling field mapping in these working devices with sensitivities of approximately 100 $(V/m)/\sqrt{Hz}$ or better, which is competitive with state of the art spin and charge based electrometry techniques (see supplemental text)(*43–46*).

**Reducing spectral diffusion using charge depletion**

Uncontrolled fluctuating electrical environments are a common problem in spin systems where they can cause dephasing(*25*), as well as in quantum emitters where they result in spectral diffusion of the optical structure and lead to large inhomogeneous broadening. For example, adding and removing just a single electron charge 100 nm away causes shifts on the order of 100 MHz for the optical fine structure of $VV^0$ (Fig. S2). Previous work(*24*) has shown that by doing an exhaustive search through many defects in a specially grown material, one can find defects with lines as narrow as 80 MHz (typically 100-200 MHz or larger), however, this is still much larger than the Fourier lifetime-limit of ~11 MHz(*24*). In bulk intrinsic commercial material, the narrowest linewidths are significantly broadened to around or above 130-200 MHz(*24*) (Fig. S3). Overall, spectral diffusion has been a notoriously difficult outstanding challenge for nearly all quantum emitters in the solid-state.

Here, we present a novel technique for mitigating spectral diffusion. We demonstrate that by applying electric fields in our device we deplete the charge environment of our defect and obtain single scan linewidths of 20±1 MHz (Fig. 3A) without the need for an exhaustive search. Using the Stark shift (Fig. 2B) to estimate the electric field, we observe this narrowing occurs with a characteristic field strength $E_0$~1.0 MV/m. This reduction in PLE linewidth has a different voltage dependence than the transverse asymmetry in the defect, thus eliminating reduced mixing as a possible mechanism for narrowing (Fig. 3B). The temperature dependence of the linewidth is roughly consistent with a $T^3$ scaling at these low temperatures(*47*) (fitted exponent 3.2±0.3, discussion in supplemental text). From the fit, a zero-temperature linewidth of 11±5 MHz is extracted (Fig. 3C). Our results are therefore consistent with a lifetime-limited line (11 MHz) that is broadened by temperature. Furthermore, the observed line is extremely stable, with a fitted inhomogeneous broadening of 31±0.4 MHz averaged for over 3 hours (Fig 3A). This stability over time, narrowness, tunability, and photostability demonstrate the effectiveness of engineering the charge environment with doped semiconductor structures for creating ideal and indistinguishable quantum emitters.

It is worth noting that at zero bias the linewidth in our samples is much higher than in bulk material (around 1 GHz, Fig. 2A). We attribute this to a greater presence of traps and free carriers (under illumination). Thus, in these samples, the observed narrowing corresponds to an improvement in the linewidth by a factor of more than 50. We speculate that a combination of this charge depletion technique with lower sample temperatures, a lower impurity material, and further annealing could enable measurement of consistent transform-limited linewidths(*13*, *48*). This use of charge depletion for creating spectrally narrow optical interfaces (Fig. 3D) could be widely applicable to other experiments in SiC, or to other solid-state emitters such as quantum dots(*49*, *50*). Indeed, by applying the same techniques developed here to intrinsic SiC materials, lines as narrow as ~21 MHz are observed(*40*). Crucially, these results demonstrate that depleting local charge environments can transform a very noisy electric environment into a clean one, turning materials containing unwanted impurities into ideal hosts for quantum emitters.

**Charge gating and photodynamics of single defects**

Our observation of large Stark shifts and linewidth narrowing relies on understanding and controlling charge dynamics under electric fields. To achieve this, we study the stability of the observed single defects under electrical bias. This allows a careful investigation of the charge dynamics of single $VV^0$ under illumination, from which we develop an efficient charge reset protocol. In our experiments, we observe that with 975 nm off-resonant light, the photoluminescence (PL) drops dramatically once a threshold voltage is reached (Fig. 4A). This threshold varies between defects, which is expected given differences in the local electric field experienced due to variations in position, depth, and local charge trap density. We attribute the PL reduction to photoionization to an optically 'dark' charge state(*18*). We use this effect to create an electrically gated single photon source(*51–53*), where emission is modulated in time with a gate voltage (Fig. 4B)(*10*). The threshold voltage has a slight hysteresis (Fig. S4) and laser power dependence (Fig. 4A) suggesting that trapped charges may play a role(*9*, *54*). We note that the electric field dependence of the photoionization could also be used to extend sensitive electrometry techniques(*46*) to the single defect regime, while controlled ionization of the spin may extend the coherence of nuclear registers(*27*). The threshold for Stark shifts (Fig. 2A) corresponds approximately to the same voltage where significant photobleaching occurs when using off-resonant excitation. This links the sharp photoionization threshold in Fig. 4A to the presence of moderate electric fields and the onset of carrier depletion.

A possible explanation for this voltage-dependent PL is that at zero electric field, illumination constantly photoionizes the $VV^0$ and other nearby traps. However, the divacancy rapidly captures available free carriers returning it to the neutral charge state. Under applied field, carrier drift depletes the illuminated region of charges. Thus, when a $VV^0$ photoionization event occurs in this depleted environment, no charges are available for fast recapture, resulting in a long-lived dark state (Fig. 4C).

Past works have shown that PL is enhanced in ensembles by repumping the charge with higher energy laser colors(*18*, *55*, *56*). We extend this work to the single defect regime by applying various illumination energies and studying single defect photodynamics past the threshold voltage (-90 V for this defect). We observe under resonant illumination the PL quickly drops to zero and does not recover, indicating that while 1131 nm (1.09 eV) light (resonant with the ZPL of a (*kk*) $VV^0$) ionizes the defect, it does not reset the charge state. However, after applying higher energy light (688 nm, for example) the charge is returned to a bright state even with <1 nW of applied power. This 'repump' of the defect charge state is vital for restoring PL for ionized or charge unstable $VV^0$ in SiC (Fig. 4A) and is essential to observe the effects discussed in the previous sections (Fig. 4C).

When both NIR resonant (1131 nm) and red (688 nm, 1.8 eV) light is applied to the defect, hopping between the bright ($VV^0$) and dark ($VV^+$ or $VV^-$) charge states results in a blinking behavior. From this blinking (Fig. S5), we can extract photoionization and repumping rates of the defect(*57*). We first examine the ionization rate of a single $VV^0$ (Fig. 5A) and observe that the power dependence is quadratic below defect saturation (exponent m=2.05±0.2) and linear at higher powers (m=0.99±0.07). We note that our observed data provide evidence for a two-photon process to $VV^-$ (see supplemental text) suggested in previous ensemble studies(*18*, *56*), while it is less consistent with a recently proposed three-photon model converting to $VV^+$ (*35*, *55*). Further study of the spin dependence of this ionization may lead to the demonstration of spin-to-charge conversion in $VV^0$.

Similarly, we study the charge reset kinetics by varying the power of the repumping laser. We find a near-linear power law with m= 0.98±0.05 (Fig. 5B). This linear dependence of the repumping rate can be described by two potential models. One possibility is that the dark charge state is directly one-photon ionized by repump laser. The other possible explanation is that nearby traps are photoionized by this color and the freed charges are captured by the divacancy to convert back to the bright state. By varying the color of this reset laser, we find repumping to be most efficient around 710 nm (1.75 eV), suggesting a particular trap state energy or a possible defect absorption resonance(*58*, *59*) (Fig. 5C). Overall, we observe negligible ionization from the optimal red repump laser and no observable reset rate from the resonant laser. This results in fully deterministic optical control of the defect charge state (discussion in supplemental text), allowing for high-fidelity charge state initialization for quantum sensing and communications protocols.

**Conclusions and outlook**

In summary, we isolate and control highly coherent single defect spin qubits in a technologically mature semiconductor device, leveraging the advantages of SiC as a host for quantum systems. Notably, the high electric fields enabled by the diode's rectifying behavior result in large Stark shifts of these quantum emitters (>850 GHz). Unexpectedly, we also demonstrate that the charge depletion region created in the device decreases the fluctuating electric noise in the defect's local environment, thus greatly narrowing the optical fine structure to linewidths approaching the lifetime limit. Combining these two results, this system displays one of the largest linewidth to tuning range ratios (>40,000 linewidths) in any single photon source which may allow for spectral multiplexing of many quantum channels(*42*). This electrical tuning of the environment constitutes a general method which could be applicable to various quantum emitters in semiconductors where spectral diffusion or charge stability is an issue(*60*), or where electric field fluctuations limit spin coherence(*25*). Furthermore, using our p-i-n diode as a testbed to study charge dynamics, we develop a technique to perform deterministic optical control of the charge state of single divacancies under electric fields

The techniques presented will be vital to achieving single-shot readout and entanglement in $VV^0$ by enabling charge control and enhancing photon indistinguishability, suggesting doped semiconductor structures as ideal quantum platforms for defects. This work also enables high-sensitivity measurement of nanoscale electric fields and charge distributions in working devices(*43*) and facilitates spin-to-charge conversion(*19*) for enhanced quantum sensing and electrical readout protocols(*20*). Finally, the introduction of $VV^0$ into classical SiC semiconductor devices such as diodes, MOSFETs and APDs, for example, may enable the next generation of quantum devices.

**Acknowledgments**

We thank E. O. Glen, S. Bayliss, D. J. Christle, P. V. Klimov, P. J. Duda for experimental suggestions and assistance and A. Gali, G. Galli, M. E. Flatté, D. R. Candido, B. Magnusson for insightful discussions and theoretical understanding. Careful reading by F. J. Heremans supported manuscript preparation. We thank *Quantum Opus* for their assistance with SNSPDs. **Funding:** This work made use of the UChicago MRSEC (NSF DMR-1420709) and Pritzker Nanofabrication Facility, which receives support from the SHyNE, a node of the NSF's National Nanotechnology Coordinated Infrastructure (NSF ECCS-1542205). C.P.A., A.B., K.C.M., G. W., P. J. M., A.L.C., and D.D.A. were supported by AFOSR FA9550-14-1-0231 and FA9550-15-1-0029, DARPA D18AC00015KK1932, NSF EFRI EFMA-1641099, and ONR N00014-17-1-3026. C.P. A. was supported by the Department of Defense through the NDSEG Program, and T.O was supported by KAKENHI (17H01056 and 18H03770). J. U. H was supported by the Swedish Energy Agency (43611-1). N. T. S. received support from the Swedish Research Council (VR 2016-04068), the Carl Tryggers Stiftelse för Vetenskaplig Forskning (CTS 15:339). J. U. H. and N. T. S. were also supported by the Knut and Alice Wallenberg Foundation (KAW 2018.0071). **Author contributions:** C.P.A and A. B conceived the experiments, fabricated the devices, performed the measurements and analyzed the data. A. B. and K. C. M developed the experimental setup. A. C. assisted in device fabrication. H. A and T. O. performed the electron irradiation. J. U. H and N. T. S. assisted in growth and sample preparation of test devices. G. W. and P. J. M. measured initial devices. D. D. A. advised on all efforts. All authors contributed to the data analysis and manuscript preparation. **Competing interests:** The authors declare no competing financial interests. **Data and materials availability:** All data are available upon request to the corresponding author.

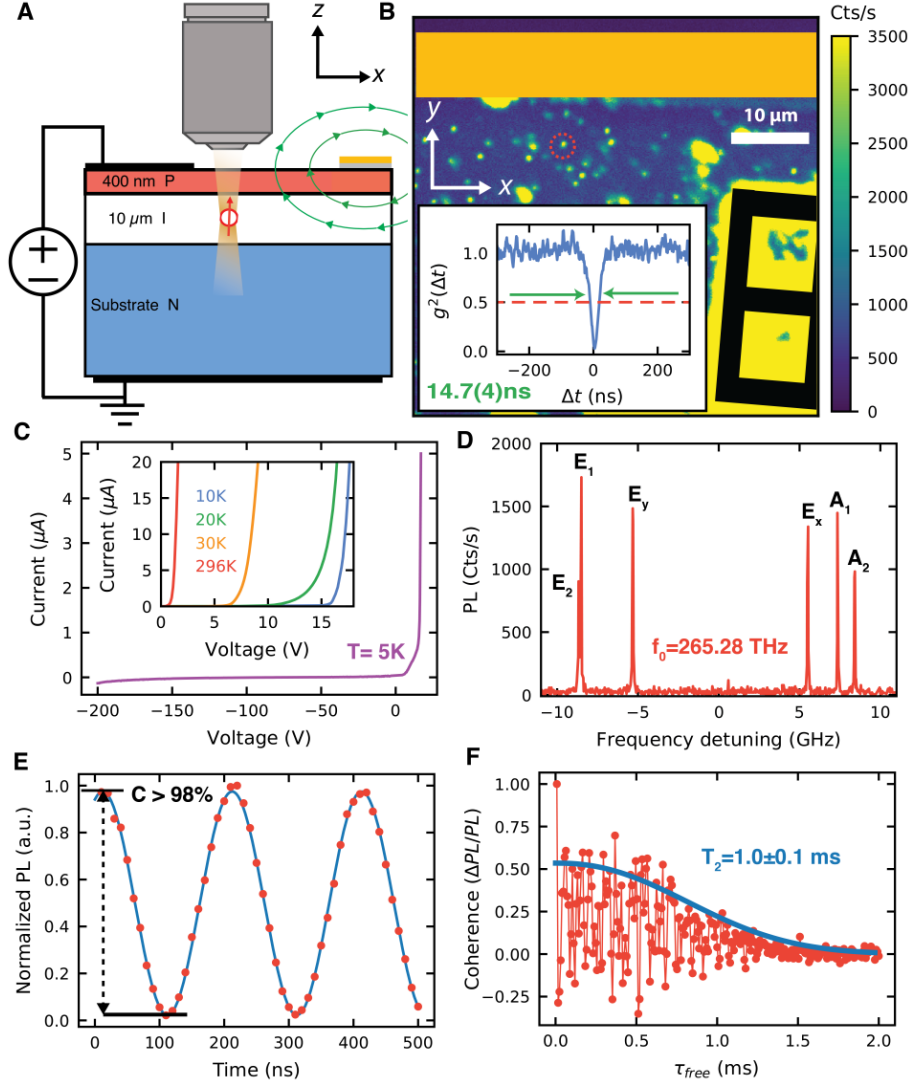

**Fig. 1. Isolation of single VV$^0$ in a commercially grown semiconductor device.** (**A**) Schematic of the device geometry. (**B**) Spatial photoluminescence (PL) scan of an example device, showing isolated emitters (example circled in red) confirmed by autocorrelation (inset) showing g$^2$(0)<0.5 (red line). Extracted emitter lifetime is 14.7±0.4 ns (green arrows). Gate and microwave stripline features are drawn and color coded as in (A). (**C**) I-V curves of the device at various temperatures. (**D**) Photoluminescence excitation (PLE) spectrum of a single (*kk*) divacancy at -270 V (**E**) Optically detected Rabi oscillations of a single (*kk*) VV$^0$ with >98% contrast (fit in blue) using resonant initialization and readout. (**F**) Hahn-echo decay of a single (*kk*) VV$^0$ in the diode. Rabi, Hahn and g$^2$ data are taken at -270 V and at approximately 240 Gauss at T=5 K.

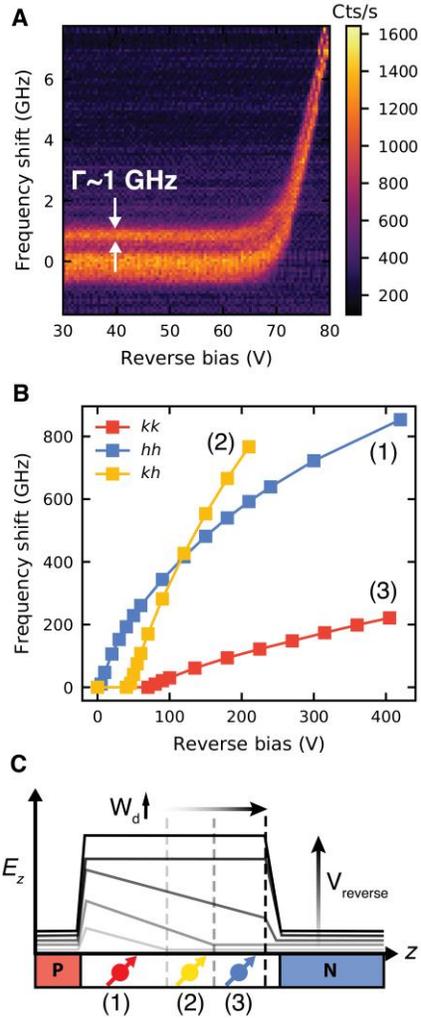

**Fig. 2. Stark shifts in p-i-n diode.** (**A**) Low field Stark tuning of a single (*kk*) defect, showing a turn-on behavior for the Stark shifts and narrowing with voltage. This threshold is the same as in Fig. 4A. These scans contain the lower branch ($E_1, E_2, E_y$) where the linewidth of $E_y$ is approximately 1 GHz and $E_1$ and $E_2$ are unresolved. The PLE lines show no shifting down to zero bias. (**B**) High field Stark shifts of multiple example defects (located at various depths and positions in the junction), showing >100 GHz shifts. (**C**) Schematic electric field distribution in the diode. Location in the junction can determine the local field experienced by the defects in (B). The electric field distribution also represents the width ($W_d$) of a depletion region in the diode. The error bars in (B) are smaller than the point size.

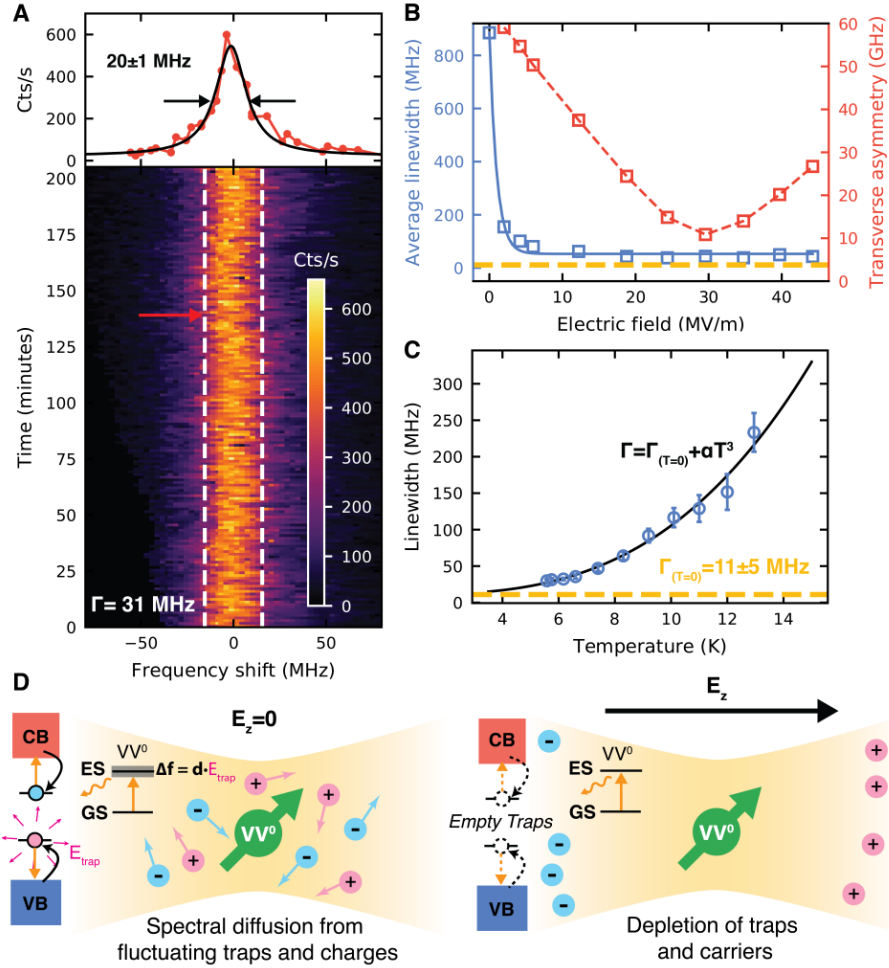

**Fig. 3. Optical linewidth narrowing by tuning the electrical environment of a solid state emitter.** **(A)** Multiple PLE sweeps taken over 3.5 hours of the $E_x$ line, showing small residual spectral diffusion (fitted inhomogeneous linewidth of 31±0.4 MHz). The red arrow corresponds to the single scan shown with a fitted linewidth of ~20 MHz. **(B)** Comparison of the average linewidth (blue) and defect transverse asymmetry (red) with respect to an estimated field (x-axis). The transverse asymmetry is defined as the differece between the $E_x$ and $E_y$ frequencies. The estimate is obtained by combining the measured Stark shifts with an estimated dipole of d=4.5 GHz m/MV for the (kk) divacancy. The yellow line is the lifetime limit. **(C)** Temperature dependence of the linewidth. A free power law fit gives an exponent of 3.2±0.3. Constraining the fit to a $T^3$ relation, we extract a zero temperature linewidth of 11±5 MHz (yellow line), consistent with the lifetime limit. Errors on the plot represent a 95% confidence interval. **(D)** Model for the effect of charge depletion on spectral diffusion in the illuminated volume (yellow). To the left of each diagram is a schematic band diagram with the relevant transitions. Errors for the fits values in (A) and (C) represent one standard deviation. All data is from a single (kk) $VV^0$. In (B), the laser power is slightly higher than in (A), causing some broadening. For (A) and (C) the $E_x$ line is shown at a voltage of -270 V (corresponding to ~30 MV/m).

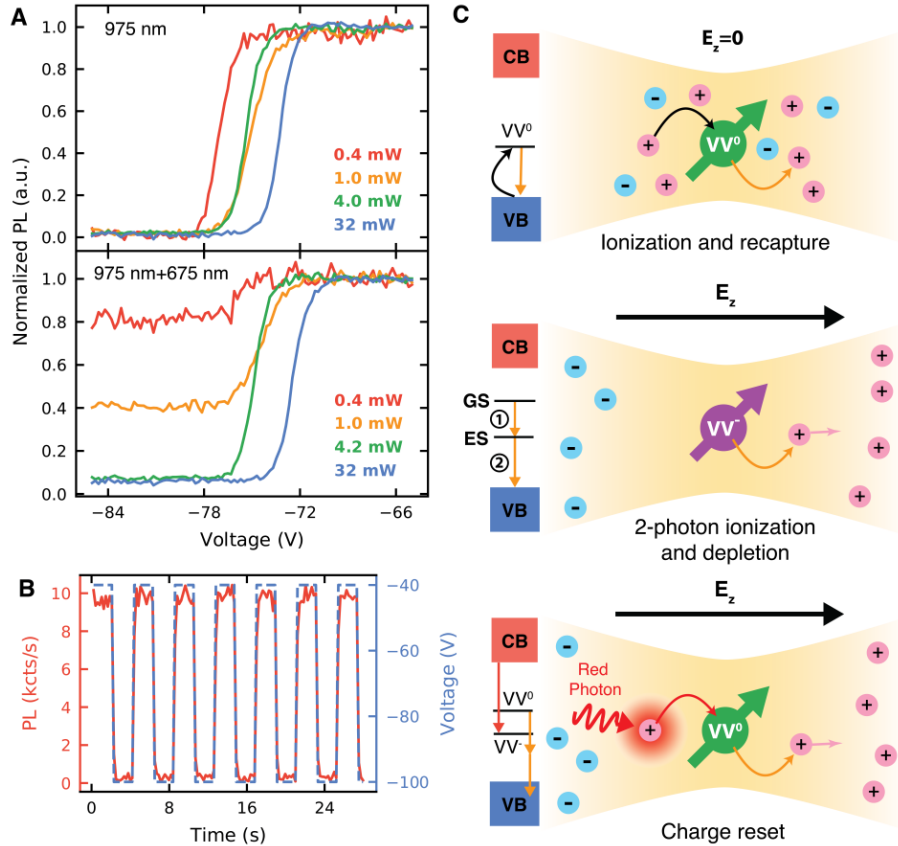

**Fig. 4. Electrical and optical charge control of a single VV$^0$.** (**A**) Voltage and power dependence of the photoluminescence of a single (*kk*) VV$^0$ with 975 nm excitation, and with 188 µW of 675 nm illumination, showing a sharp threshold under reverse bias. With high 975 nm power, the two-photon ionization process dominates and the PL signal is low. (**B**) By controlling the voltage in time (blue) the emission from the single (*kk*) defect is switched on and off (red). (**C**) Model of rapid ionization and recapture at zero electric field (top). Two photon ionization and formation of a depletion region under reverse bias (middle). Charge reset under applied electric field using red light (bottom).

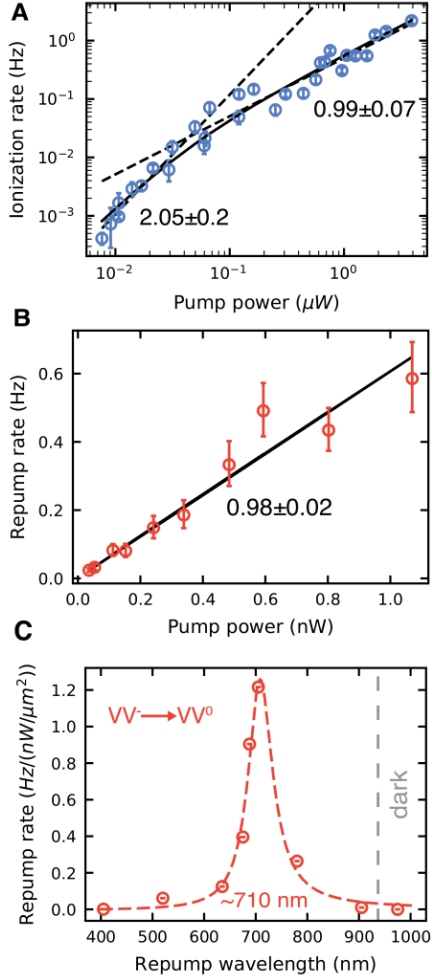

**Fig. 5. Ionization and charge reset rates for VV$^0$.** (**A**) Dependence of the ionization rate on resonant laser power. Low and high power regime fits (black dotted lines) and their power laws (m=2.05±0.2 and m=0.99±0.07, respectively). The solid black line shows a full model fit. (**B**) Repump power dependence of the 688 nm laser, showing a linear exponent of m=0.98±0.2. Fluctuations in the polarization or power of the laser may limit the true error. (A) and (B) were taken at -90 V bias. (**C**) Repumping rate as a function of illumination wavelength at -270 V with a Lorentzian fit centered around 710 nm. Higher than 905 nm (and at these powers) no PL is observed. All error bars represent 95% confidence intervals from the fit of the raw data from a single (*kk*) VV$^0$.